\documentclass[aps,twoside,showpacs,superscriptaddress,twocolumn,floatfix,a4paper]{revtex4}

\usepackage{color}
\usepackage{graphicx}
\usepackage[utf8x]{inputenc}
\usepackage[T1]{fontenc}

\newcommand\ket[1]{|#1\rangle} 
\newcommand\bra[1]{\langle #1|}
\newcommand\oprod[2]{\ket{#1}\bra{#2}}

\begin{document}

\title{Entanglement-assisted scheme for nondemolition detection\\ of the presence of a single photon}

\author{Marek Bula}
\affiliation{RCPTM, Joint Laboratory of Optics of Palacký University and Institute of Physics of Academy of Sciences of the Czech Republic, 17. listopadu 12, 772 07 Olomouc, Czech Republic}
\author{Karol Bartkiewicz}
\email{bartkiewicz@jointlab.upol.cz}
\affiliation{RCPTM, Joint Laboratory of Optics of Palacký University and Institute of Physics of Academy of Sciences of the Czech Republic, 17. listopadu 12, 772 07 Olomouc, Czech Republic}
\author{Antonín Černoch}
\affiliation{Institute of Physics of Academy of Sciences of the Czech Republic, Joint Laboratory of Optics of PU and IP AS CR, 
   17. listopadu 50A, 779 07 Olomouc, Czech Republic}
\author{Karel Lemr}
\email{k.lemr@upol.cz}
\affiliation{RCPTM, Joint Laboratory of Optics of Palacký University and Institute of Physics of Academy of Sciences of the Czech Republic, 17. listopadu 12, 772 07 Olomouc, Czech Republic}

\date{\today}

\begin{abstract}
In this paper, we propose a resources-optimal linear-optical scheme for quantum nondemolition detection of single-photon presence. By measuring the state of ancillary photons, the presence of a photon in signal mode is revealed with a success probability of 1/2 without any disturbance to its state. We also show how to tune the setup to perform quantum nondemolition measurement of the signal photon state, and we provide tradeoff between the extracted information and the signal state disturbance.  Moreover, the optimality of resources and methods by which to increase the success probability are discussed. 
\end{abstract}


\maketitle

\section{Introduction}
The platform of linear optics is a prominent platform suitable for quantum information processing \cite{bib:kok}. Its major advantage is high experimental accessibility and relatively low cost \cite{bib:kiesel}. Hence many of the most important quantum information protocols, like teleportation \cite{bib:zeilinger}, cloning \cite{bib:buzek, bib:cernoch, bib:sciarrino, bib:lemr}, and various quantum gates \cite{bib:brien,bib:mikova,bib:lemr02, bib:kiesel, bib:cernoch02}, have been demonstrated using individual photons and linear-optical components. On the other hand, linear-optical quantum computing is burdened by two significant drawbacks related to its probabilistic nature \cite{bib:klm, bib:knill, bib:probabilisticManipulation}: first, the success probability decreases exponentially with the number of quantum gates, and second, in a large number of cases there is a need for postselection in order to distinguish the successful and unsuccessful operations of these probabilistic gates. There are several proposals designed to circumvent the first mentioned issue by employing high-photon-number ancillary states \cite{bib:klm, bib:cluster, bib:chuang}. The second issue can be circumvented using quantum nondemolition measurement \cite{bib:grangier}. 

To illustrate the role of quantum nondemolition measurement in linear-optical schemes, let us consider an example of optimal implementation of the controlled-NOT (CNOT) gate, assuming no additional photon ancillae. Such device is successful only in one in nine cases \cite{bib:kiesel} (for other implementations see review \cite{bib:bartkowiak}). These cases correspond to both the signal and control qubits leaving the gate by their respective output ports. If only one such  gate is used, the simple postselection on coincidence detection in signal and control modes suffices. If, however, one wishes to chain several gates of this kind, one has to be able to detect the presence of photons in signal and control modes in between the gates without directly measuring them and thus disturbing their quantum states. At this point the quantum nondemolition measurement looks particularly suitable. Moreover, as it has been recently discussed in \cite{bib:eavesdropping} the nondemolition photon detection can be applied for triggering a cloning machine set for eavesdropping on a quantum key distribution protocol. The operator of any cloning machine used in a real life situation has to synchronize the generation of the ancillary photon with the incoming signal photon. Since, for instance in quantum cryptography, the timing of signal photons is unknown to the cloner operator, the nondemolition presence detection will be a crucial element of a practical cloning attack.

Several schemes for linear-optical quantum nondemolition measurement (QND) of individual photons have been proposed and experimentally verified. For instance, a  CNOT gate can be used to investigate the state of a qubit without destroying it by detection \cite{bib:pryde04,bib:pryde05, bib:pryde06,bib:pryde10}. This approach utilizes the conditional SWAP implemented by the gate on the signal qubit when the control qubit is in a specific state. Thus observing such a SWAP gate on the signal qubit allows to determine the state of the control qubit without its direct detection, which would have destroyed it. The optimal average success probability of this scheme is found to be $\frac{1}{6}$. An alternative strategy is based on using polarizing beam splitter and feed-forward \cite{bib:filip06a, bib:filip06b, bib:filip12, bib:barbieri}. This strategy increases the success probability of the QND measurement to $\frac{1}{2}$, and it has also been used to test the boundaries of extracted information versus disturbance of the measured state \cite{bib:filip05}. Apart from linear-optical schemes, the quantum nondemolition measurement can also be implemented using nonlinear optical interaction \cite{bib:kapale}. This approach is however limited by the amount of nonlinearity in currently available media.

Papers mentioned in the preceding paragraph discuss the QND measurement of the information encoded into the signal photon state, while in this paper we address a somewhat different problem: the goal is to detect the mere presence of a photon in the signal spatial mode without disturbing the information encoded into its polarization state. Such a measurement can in principle be described by the following unitary transformation:
\begin{equation}
\hat{U} = \oprod{0}{0}\otimes\openone_2 +  \oprod{1}{1}\otimes\hat{V}
\end{equation}
 where $\oprod{0}{0}$ correspond to the vacuum state of the investigated 
input and $\oprod{1}{1}$ corresponds to the single photon that we would like to detect without 
disturbing it. The second subsystem is the ancillary system, which is unchanged if there is a vacuum state in the signal mode but undergoes transformation $\hat{V}$ if there is a signal photon. The transformation $\hat{V}$ is selected so that the change on the ancillary subsystem can be deterministically observed. Capability of performing this type of presence detection is crucial for combining several postselection based quantum gates into larger circuits. Later in Section III we also show how to set the setup to perform a tunable QND measurement on the signal qubit at the expense of its partial disturbance.

\section{Principle of operation}
\begin{figure}
\includegraphics[scale=0.29]{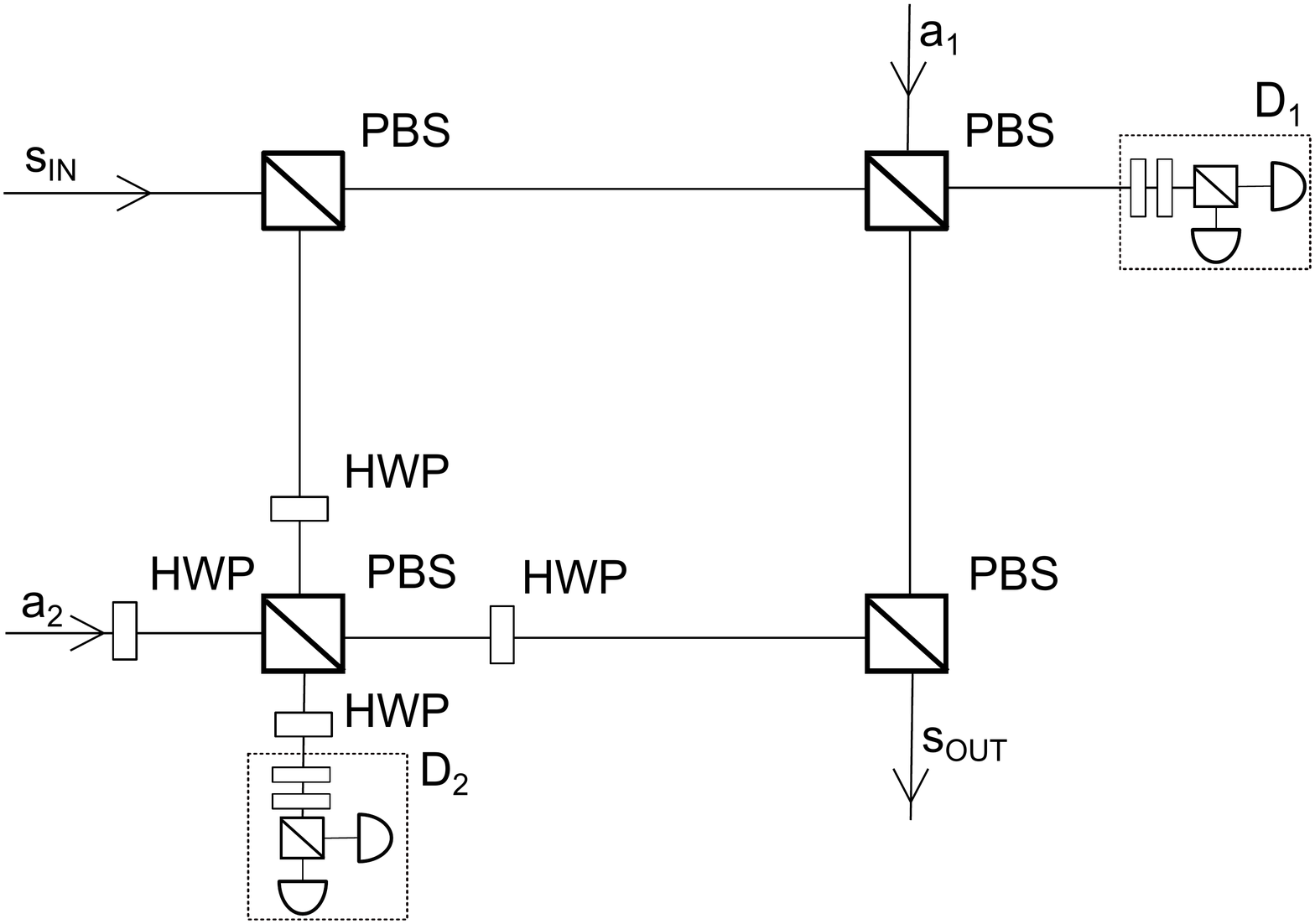}
\caption{\label{fig:scheme} Proposed setup for linear-optical quantum nondemolition measurement of single photon; PBS, polarizing beam splitter; HWP, half wave plate (all rotated by 45 deg. with respect to horizontal polarization direction); s, signal photon mode; $a_1$, $a_2$--ancillary photon modes; $D_1$ and $D_2$, standard polarization analysis (for reference see \cite{bib:twoPhotonAnalyzer}).}
\end{figure}
The proposed scheme for linear-optical quantum nondemolition measurement is depicted in Fig.~\ref{fig:scheme}. It consists of four polarizing beam splitters (PBSs) transmitting horizontally polarized light and reflecting vertically polarized light,  four half wave plates (HWP) set to perform horizontal-vertical polarization swap ($H \to V$) and two detectors $D_1$ and $D_2$. Apart from the signal photon entering the device by input port $s_\mathrm{in}$ there are also two entangled ancillary photons entering by input ports $a_1$ and $a_2$. Successful nondemolition detection of the signal photon is obtained when two-photon coincidence detection on the detectors $D_1$ and $D_2$ is observed. As derived below, this occurs with a probability of $\frac{1}{2}$  when the signal photon is present and with a zero rate if there is no signal photon.

Let us assume the signal photon entering the scheme in an arbitrary polarization state:
\begin{equation}
\label{eq:signal}
|\psi_s\rangle = \alpha \ket{H} + \beta \ket{V}, \\
\end{equation}
where $|H\rangle$ and $|V\rangle$ denote horizontal and vertical polarization states, respectively, and the coefficients follow normalization condition $|\alpha |^2 + |\beta |^2 =1$. The ancillary photons are initially in a maximally entangled Bell state
\begin{equation}
\ket{\psi _{a_1a_2}} = \frac{1}{\sqrt{2}}(|HH\rangle + |VV\rangle)_{a_1,a_2}.
\end{equation}
Note that preparation of such an entangled state is nowadays a routine procedure\cite{bib:kwiat99}.
One can therefore formulate the total three-photon state entering the aparatus:
\begin{eqnarray}
\label{Eq:TotalState} 
|\psi_{T}\rangle & = & \frac{1}{\sqrt{2}}(\alpha \ket{HHH} + \alpha \ket{HVV} \\ \nonumber 
&&+ \beta \ket{VHH} + \beta \ket{VVV})_{s,a_1,a_2},
\end{eqnarray}
where the order of photons is signal ($s$), first ancillary ($a_1$) and second ancillary ($a_2$) photon.  

In order to understand the transformation that the total three-photon state undergoes in the setup, we need to study the transformation of its respective components. As one can easily verify the state $\ket{HHH}$ goes through the scheme unchanged and leads to two photons impinging the detectors $D_1$ and $D_2$, while one photon with horizontal polarization leaves the scheme by the signal output port $s_\mathrm{out}$. Similarly the state $\ket{VVV}$ always passes through the device unmodified leading as well to two-photon coincidence on the detectors and vertically polarized photon leaving the device.

On the other hand, one can observe that in the two remaining cases ($|HVV\rangle$ and $|VHH\rangle$) the states never lead to two-photon coincidence on the detectors $D_1$ and $D_2$. In these cases only one of the detectors registers a detection event. These cases are excluded from the output state by coincidence postselection. Since the probability of such an outcome is $\frac{1}{2}$, the overall success probability of the scheme is the remaining $\frac{1}{2}$.

Taking into account the transformation of the input state and the postselection on detection coincidences, the total state at the output before detection reads
\begin{equation}
|\psi_T\rangle=(\alpha |HHH\rangle +\beta |VVV\rangle)_{s,a_1,a_2},
\label{PsiOut}
\end{equation}
where renormalization has been carried out having the success probability of $\frac{1}{2}$  in mind. We now perform a projection polarization measurement in the output ancillary ports using diagonal $|D\rangle$ and antidiagonal $|A\rangle$ linear polarization basis. Using this basis, one can rewrite the total output state as
\begin{eqnarray} 
|\psi_T\rangle &=&\frac{1}{2}[\alpha|H\rangle(|DD\rangle+|AA\rangle +|DA\rangle+|AD\rangle)\\ \nonumber
&&+\beta|V\rangle(|DD\rangle+|AA\rangle- 
|DA\rangle-|AD\rangle]_{s,D_1,D_2},
\end{eqnarray}  
where indexes $D_1$ and $D_2$ denote photons impinging on the respective detectors. If $|DD\rangle$ or $|AA\rangle$ coincidence is detected on the ancillary photons the signal photon (in the signal output mode) is projected directly into its initial state (\ref{eq:signal}). If $|AD\rangle$ or $|DA\rangle$ coincidences are observed the signal photon is projected into the state
\begin{equation}
|\psi_\mathrm{out}\rangle = \alpha |H\rangle - \beta |V\rangle, \\
\label{Eq:SignalMinus}
\end{equation}
which can be easily reverted to the initial state Eq.~(\ref{eq:signal}) by inserting a half-wave plate with optical axis coinciding with vertical polarization direction to the signal output port and thus implementing the transformation $V\rightarrow -V$ on the signal photon. 
If one wishes to avoid such feed-forward correction , the gate can still be operated but with a reduced success probability of $\frac{1}{4}$. since $|AD\rangle$ and $|DA\rangle$ measure outcomes have to be neglected. Note that the particular choice of the measurement basis leads to restoration of the signal state to its exact initial form, while giving no information about its polarization state. Only the presence of the signal photon is witnessed.

To complete the derivation of the principle of operation, let us consider the case in which there is no signal photon. Such total state reads
\begin{equation}
|\psi_T\rangle = \frac{1}{\sqrt{2}}\left(|0HH\rangle + |0VV\rangle\right),
\end{equation}
where $|0\rangle$ denotes the absence of signal photon. It is easy to observe that neither $|0HH\rangle$ nor $|0VV\rangle$ can lead to a coincidence on detectors $D_1$ and $D_2$. Therefore, observing such coincidence can only happen when the signal photon is present.

\section{Tunable weak measurement}
Apart from the single photon presence detection, the scheme can also be used to measure the polarization state of the signal photon in a nondemolition manner. Equation~(\ref{PsiOut}) indicates perfect correlation between the polarizations of the signal and the ancillary photons and therefore the measurement on ancillary photons reveals the polarization of the signal photon. Such measurement would, however, disturb the initial signal state by projecting it depending on the measurement outcome. In the previous section, this correlation has been intentionally erased by choosing suitable detection basis to prevent the disturbance of the signal state.

The scheme can provide tunable strength of the correlation between the signal and ancillary photons and thus enable also tunable quantum nondemolition measurement. This is achieved simply by rotation of the ancillary photons measurement basis. Let us denote the basis rotation angle by $\phi$ and express the transformation of the basis explicitly
$$
        \left(
                \begin {array}{c}
                \ket{H} \\
                \ket{V}
                \end {array} 
        \right)
              \rightarrow 
        \left(
                \begin {array} {cc}
                  \cos\phi & \sin\phi \\
                   -\sin\phi & \cos\phi
                \end {array}
        \right) 
        \left(
                \begin {array}{c}
                \ket{H} \\
                \ket{V}
                \end {array} 
        \right).  
$$     
Substituting the new basis transformation to the original Eq.~(\ref{PsiOut}) gives the overall state of the system in the form of
\begin{eqnarray} 
\label{eq:out_phi}
|\psi_T\rangle &=& \lbrace\alpha[\cos^2\phi\ket{HHH}+\cos\phi\sin\phi\ket{HHV} \\ \nonumber
&& +\cos\phi\sin\phi\ket{HVH}+\sin^2\phi\ket{HVV}]\\\nonumber
&&+\beta[\cos^2\phi\ket{VVV}-\cos\phi\sin\phi\ket{VHV} \\ \nonumber 
&&-\cos\phi\sin\phi\ket{VVH}+\sin^2\phi\ket{VHH}]\rbrace_{s,a_1,a_2},
\end{eqnarray} 
where the photons are ordered as usual in the order of signal, first ancillary, second ancillary. Note that subsequent calculations assume that similarly to Eq.~(\ref{Eq:SignalMinus}) the $V \to -V$ transformation is performed on the signal photon whenever the ancillary photons are detected having mutually orthogonal polarizations.  
Hence, we observe tunable correlation between the signal and ancillary photon polarizations and consequently also tunable degree of disturbance of the signal state after the ancillae are measured. This leads to a tradeoff between the obtained polarization information and the signal state disturbance.

There are several approaches to quantify such a tradeoff depending on the way we calculate the correlation between the signal and ancillary photons. First, let us employ the coherent information $I_c$ \cite{bib:quatumData}. This correlation measure is defined by means of the von Neumann entropy
\begin{equation}
\label{eq:iq}
I_c \equiv S(\hat\rho_\mathrm{out}) - S(\hat\rho_T),
\end{equation}
where $S(\hat\rho_\mathrm{out})$ is the von Neumann entropy of the subsystem of the signal photon [$\hat\rho_\mathrm{out}$ is obtained from Eq.~\ref{eq:out_phi}] by tracing over the ancillary photons) and $S(\hat\rho_T)$ is the von Neumann entropy of the entire system. Since the entire system is in a pure state the last term in Eq.~ (\ref{eq:iq}) vanishes. Moreover, since we deal with pure states the coherent information is equivalent to relative entropy of entanglement \cite{bib:quantifyingEntanglement} and quantifies how much the output state is entangled with the apparatus. In the specific case of our scheme, the coherent information can be easily expressed in terms of the signal state parameters $\alpha$ and $\beta$ and the basis rotation angle $\phi$
\begin{equation}
\label{eq:entr_ent_param}
I_c = - \sum_{i=1,2} \lambda_i \log_2 \lambda_i,
\end{equation}
where $\lambda_{1,2} = \frac{1}{2}(1\pm\sqrt{D})$ with
\begin{equation}
D = 1-4|\alpha|^2 |\beta|^2 (1-\sin^4 2\phi).
\end{equation}
Since $\lambda_1 + \lambda_2 \equiv 1$, the coherent information $I_c$ is also equal to the binary entropy \cite{bib:quatumComputation}. 

\begin{figure}
\includegraphics[scale=1]{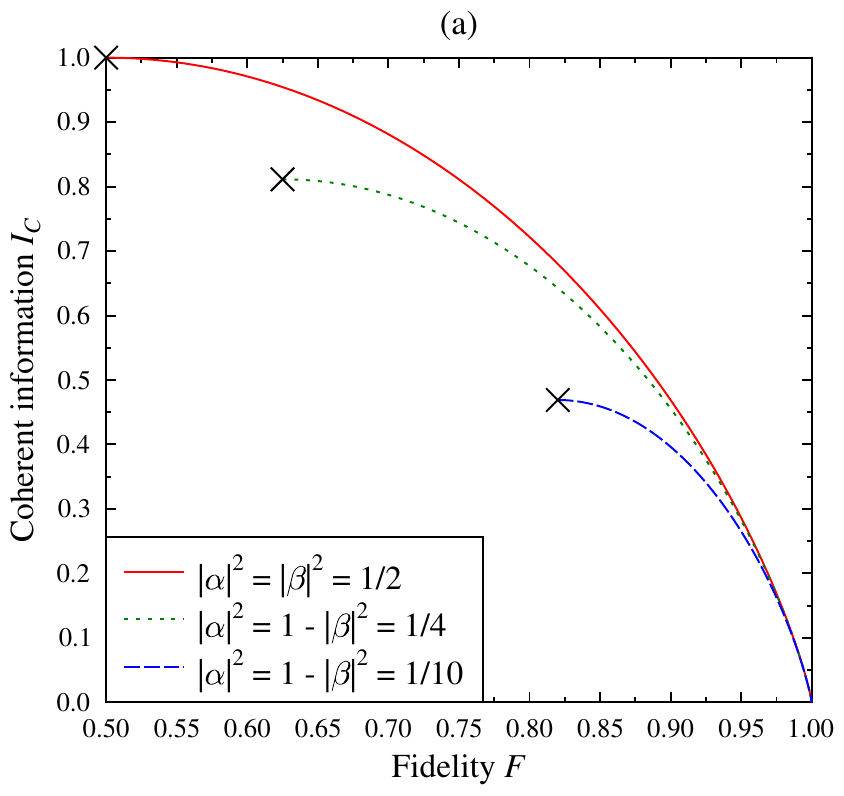}\\
\includegraphics[scale=1]{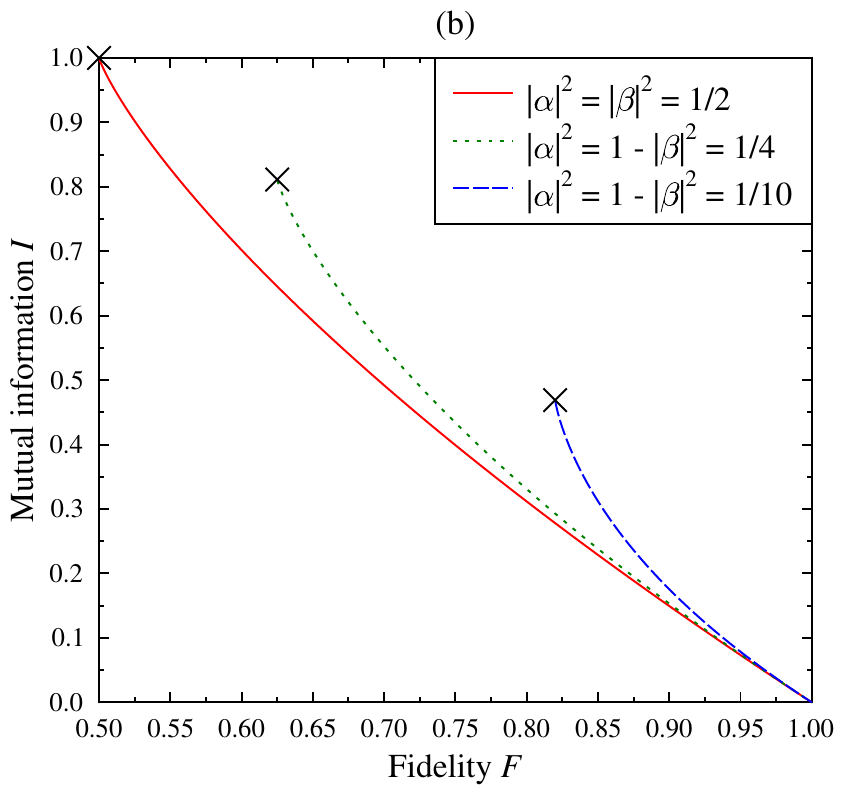}
\caption{\label{fig:plot_info} (Color online) (a) 
Output state fidelity $F$ depicted as a function of (a) coherent information $I_c$, (b) mutual information $I$ between probe and signal, both for three different input states with parameters  $\alpha$ and $\beta$. Note that the maxima in all cases coincide with the maximum physically obtainable information (Holevo bound, depicted by black crosses).}
\end{figure}

Inevitably, the capability to obtain some information about the signal state is accompanied by its partial disturbance. We measure such disturbance by the output state fidelity defined as
\begin{equation}
F = \bra{\psi_s} \hat{\rho}_\mathrm{out} \ket{\psi_s}, 
\end{equation}
where $\hat{\rho}_\mathrm{out}$ denotes the generally mixed state of the signal photon at the output obtained as described above and $\ket{\psi_s}$ stands for the initial signal photon state. 
We can explicitly find the formula for fidelity as a function of $\alpha$, $\beta$, and $\phi$:
\begin{equation}
\label{eq:fidelity}
F=|\alpha|^4 + |\beta|^4 + 2|\alpha|^2 |\beta|^2 \sin^22\phi.
\end{equation}
Now we plot the tradeoff between the coherent information $I_c$ and the fidelity $F$ [see Fig. \ref{fig:plot_info}(a)]. This plot shows that the maximum of coherent information coincides with the Holevo bound \cite{bib:quatumComputation} indicating that no more information can be extracted from the signal state by any measurement {or in terms of channel capacity -- how much information can be transmitted through our device while sending a single qubit}. Such a maximum is obtained for all three input states for $\phi = 0$, and the corresponding fidelity reaches its minimum at this point.
%

From the operational point of view, the classical counterpart of coherent information -- the mutual information $I$ -- is more suitable \cite{bib:quatumComputation}. It quantifies the amount of obtained information about the signal state by a specific measurement performed on the state of the first ancillary photon. The mutual information $I$ is defined as
$$
I = \sum_{i,j} P_{i,j}  \log_2 \frac{P_{i,j}}{P_i P_j},
$$  
where $P_i$ is the marginal probability of the signal photon having a given polarization (horizontal or vertical), $P_j$ is the marginal probability of the first ancillary photon having given polarization and $P_{i,j}$ is the joint probability of  both photons having given polarizations. It is evident that $I$ depends on the signal state parameters $\alpha$ and $\beta$ and the angle of rotation $\phi$. One can calculate the explicit formula for the mutual information:
\begin{eqnarray}
\label{eq:mutual_info}
I&=&|\alpha|^2 \cos^2\phi \log_2 \frac{\cos^2\phi}{|\alpha|^2 \cos^2\phi+|\beta|^2 \sin^2\phi} \\\nonumber
&& +|\alpha|^2 \sin^2\phi \log_2 \frac{\sin^2\phi}{|\alpha|^2 \sin^2\phi+|\beta|^2 \cos^2\phi} \\\nonumber
&&+ |\beta|^2 \cos^2\phi \log_2 \frac{\cos^2\phi}{|\beta|^2 \cos^2\phi+|\alpha|^2 \sin^2\phi}\\ \nonumber
&&+|\beta|^2 \sin^2\phi \log_2 \frac{\sin^2\phi}{|\beta|^2 \sin^2\phi+|\alpha|^2 \cos^2\phi}.
\end{eqnarray}
Again, we plot the tradeoff function this time between the fidelity $F$ and the mutual information $I$ for three different input states [see Fig. \ref{fig:plot_info}(b)]. This plot also illustrates that similarly to coherent information, the maximum of mutual information also coincides with the Holevo bound.

\begin{figure}
\includegraphics[scale=1]{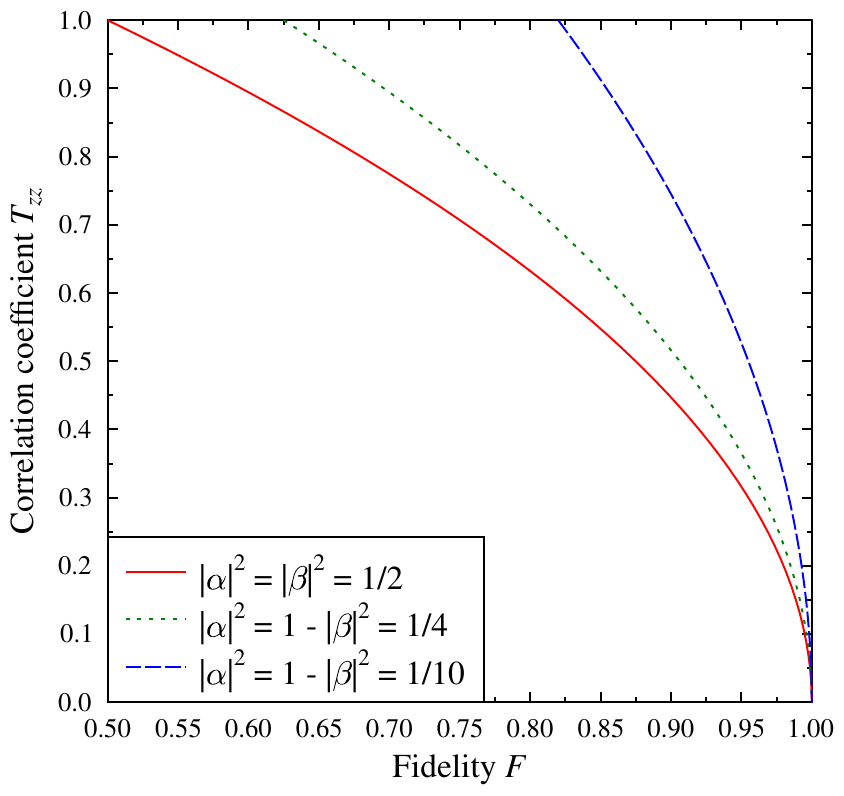}
\caption{\label{fig:plot_correl} (Color online) Output state fidelity $F$ vs correlation coefficient $T_{zz}$ (correlation between signal and probe measurements in the $H,V$ polarization basis)  depicted for three different input states with parameters $\alpha$ and $\beta$.}
\end{figure}

Both the coherent and mutual informations are well suitable for quantifying the correlation between the signal and ancillary photons. The only drawback of these measures lies in the dependence on input state parameters. For instance maxima of both these information quantities are functions of the signal state parameters $\alpha$ and $\beta$, which makes direct comparison across different input state difficult. Let us therefore use another correlation measure, this time a state independent one. A suitable candidate for such measure is the
correlation matrix $T$ of the signal and ancillary photon
polarizations 
\begin{equation}
T_{mn}=\langle\hat{\sigma}_m\otimes\hat{\sigma}_n\otimes\openone_2 \rangle_T,
\end{equation}
where $\hat{\sigma}_n$ for $n = x,\,y,\,z$ are Pauli  matrices and the subscript $T$ denotes $\ket{\psi_T}$.  Knowing the elements of this tensor it is easy to check \cite{bib:bellInequality, bib:bellInequality2} if the correlations violate Bell-Clauser-Horne-Shimony-Holt (CHSH) inequalities \cite{bib:einsteinPodolskyParadox, bib:testLocalHidVariables}. However, this goes beyond the scope of our paper. The tensor $T_{mn}$ can be expressed using joint probabilities of particular measurement outcomes on the signal and ancillary photons. For the purposes of this paper, we study only the $T_{zz}$ component of the correlation tensor defined as
\begin{equation}
\label{eq:correl}
T_{zz} = P_{H,H}+P_{V,V}-P_{H,V}-P_{V,H},
\end{equation}
where $P_{i,j}$ denotes the probability of observing the signal photon in $i$ polarization state while the first ancillary photon is found to be $j$ polarized for $i$ and $j$ being the horizontal or vertical polarizations. From Eq.~(\ref{eq:out_phi}) it follows that the correlation coefficient can be expressed  in terms of the the basis rotation angle $\phi$:
\begin{equation}
\label{eq:correl_phi}
T_{zz} = \cos 2\phi.
\end{equation}
The observed $T_{zz}$  coefficient correlation--fidelity tradeoff is illustrated in Fig. \ref{fig:plot_correl} on a set of three initial signal state.

\section{Deterministic mode of operation}
In Sec. II, we have determined that the success probability of detecting the presence of the signal photon is $\frac{1}{2}$. The successful cases are distinguished by observing coincidence detection on both detectors $D_1$ and $D_2$. For completeness, let us consider the remaining half of the cases, when two photons impinge on one detector (either $D_1$ or $D_2$). This corresponds to the second and the third term in Eq.~(\ref{Eq:TotalState}). The second term $|HVV\rangle$ leads to detecting two photons on the detector $D_1$ and the third photon leaves the scheme by the signal output mode having vertical polarization. Similarly, the third term $|VHH\rangle$ leads to two-photon detection on the second detector and, a horizontally polarized photon leaves by the output port. According to Eq.~(\ref{Eq:TotalState}) these two cases happen with probabilities $|\alpha|^2/2$ and $|\beta|^2/2$, respectively. Even though the output signal state does not equal the initial state, the output photon can still be used at the expense of lower average fidelity with respect to the initial state. In order to maximize the average fidelity, we insert a half-wave plate rotated by 45deg to the output mode when either one of the two above-mentioned cases is observed. This way, one finds a horizontally polarized photon at the output port with probability $|\alpha|^2/2$ and a vertically polarized output photon with probability $|\beta|^2/2$. The average fidelity of the output state calculated by averaging over all possible input states and taking into account all ancillary photon detection outcomes is found to be $\frac{5}{6}$. Since now there is no postselection on a particular class of outcomes, the success probability of the scheme is 1.

\section{Optimal resources for linear-optical non-disturbing photon detection} 
 
\begin{figure}
\includegraphics[width=4cm]{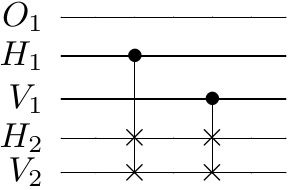}
\caption{\label{fig:circuit} Quantum-logical circuit explaining the operation of 
the non-perturbing measurement: $O_1$,  vacuum; $H_i$ 
($V_i$),  single $H$($V$)-polarized photon in $i$th spatial mode, $i=1$ denotes 
the analyzed photon and $i=2$ stands for the ancillary photon. Note that 
implementing the CNOT  gate with linear optics without using measurement-induced
nonlinearity is impossible; hence, each of such gates requires 
at least one measurement, where the last measurement could correspond to 
$\hat{\sigma}_z$ of the ancillary qubit.}
\end{figure}

In this section,  we argue that using two  ancillary photons for QND 
measurement  is optimal for its linear-optical implementation. 
By performing the QND we would like to detect the presence of a photon without disturbing its state. 
This is achieved  by measuring the $\hat\sigma_z$  
observable for the ancillary system which can be initially found in the 
$\ket{H}$ state, can be described by the following unitary transformation
\begin{equation}
\label{eq:U}
\hat{U} = \oprod{0}{0}\otimes\openone_2 +  (\oprod{H}{H} +\oprod{V}
{V})\otimes\hat{\sigma}_x
\end{equation}
 where $\oprod{0}{0}$ correspond to the vacuum state of the investigated 
input and $\oprod{H}{H}$ or $\oprod{V}{V}$ denote $\hat{\sigma}_z$ 
eigenstates of the single photon that we would like to detect without 
disturbing it. The transformation flips the ancillary state
if there is a photon to be detected and does nothing if there is none.  The 
transformation given by Eq.~(\ref{eq:U}) can be described by means
of a quantum logical circuit shown, in the Fig.~\ref{fig:circuit}.  Since $H$ and 
$V$ modes are orthogonal, we need at least two controlled
SWAP gates (or equivalently CNOT gates). Using linear optics one can 
implement nonlinear gates, i.e., nontrivial two-photon gates, by exploiting 
measurement-induced nonlinearity (for review see 
Ref.~\cite{bib:bartkowiak}). Thus, one needs  at least two ancillary photons, 
which need to be absorbed in order to perform the discussed type of the QND 
measurement. The first one is the target photon of the QND interaction 
(modes $H_2$ and $V_2$ in Fig.~\ref{fig:circuit}) and the second one ensures 
that the first of  CNOT gates operates in the nondestructive fashion.
Hence, our linear-optical implementation meets the requirements for the 
minimal number of photons involved in the QND interaction.
Moreover, we propose exploiting the entanglement of the ancillae to enhance 
the probability of the QND operation with respect to the consecutive usage of 
the best know nondestructive optical CNOT gates (success rate of a single 
operation is $p = \frac{1}{4}$). Nevertheless, as we discuss in the section,
implementation of the non-perturbing detection in a nonlinear system
requires far less resources. 
 
\section{Conclusions}

In this paper, we have proposed a resource efficient scheme for nondemolition detection of single photon presence. By performing measurement on a pair of ancillary photons, one can detect the presence of a single photon in signal mode. Such detection does not disturb the information encoded into polarization state of the signal photon and can therefore be used as nondemolition postselection in between probabilistic gates.

We have also shown how to tune the setup to perform a QND measurement on the signal polarization state. The strength of this measurement can be tuned by simple choice of the projection performed on the ancillary photons. In order to characterize the strength of this QND measurement and its impact on the signal state, we have provided three tradeoff curves plotting output state fidelity against coherent information, mutual information and the appropriate correlation coefficient obtained by ancillary photons projection.

The original proposal is limited to $\frac{1}{2}$ success probability (or $\frac{1}{4}$ without feed-forward). We also add a simple analysis indicating that this success probability can be increased to 1 at the expense of lower output state fidelity.

\section{Acknowledgement}

The authors thank Adam Miranowicz and Jan Soubusta for fruitful discussions on the subject of this paper. The authors gratefully acknowledge the support by the Operational Program Research and Development for Innovations -- European Regional Development Fund (project No. CZ.1.05/2.1.00/03.0058)  and the  Operational Program Education for Competitiveness - European Social Fund (projects Nos. CZ.1.07/2.3.00/20.0017, CZ.1.07/2.3,00/20,0058, CZ.1.07/2.3.00/30.0004 and CZ.1.07/2.3.00/30.0041) of the Ministry of Education, Youth and Sports of the Czech Republic. The authors also acknowledge the support by the Internal Grant Agency of Palacky University in Olomouc (project No. PrF\_2012\_003) and by Czech Science Foundation (project No. P205/12/0382). K.B. was supported by Grant No. DEC-2011/03/B/ST2/01903 of the Polish National Science Centre.


\end{document}